# Wafer-scale, full-coverage, acoustic self-limiting assembly of particles on flexible substrates

Liang Zhao[1,2], Bchara Sidnawi[1,3#], Jichao Fan[4#], Ruiyang Chen[4], Thomas Scully[1,2], Scott Dietrich[5], Weilu Gao[4], Qianhong Wu[1,3], Bo Li[1,2,*]

[1] Department of Mechanical Engineering, Villanova University, Villanova, PA 19085, USA.

[2] Hybrid Nano-Architectures and Advanced Manufacturing Laboratory, Villanova University, Villanova, PA, 19085, USA

[3] Cellular Biomechanics and Sports Science Laboratory, Villanova University, Villanova, PA, 19085, USA

[4] Department of Electrical and Computer Engineering, University of Utah, Salt Lake City UT, 84112, USA

[5] Department of Physics, Villanova University, Villanova, PA, 19085, USA

[#] These authors contributed equally to this work.

[*] Corresponding author: bo.li@villanova.edu

**Self-limiting assembly of particles represents the state-of-the-art controllability in nanomanufacturing processes where the assembly stops at a designated stage[1,2], providing a desirable platform for applications requiring delicate thickness control[3-5]. Most successes in self-limiting assembly are limited to self-assembled monolayers (SAM) of small molecules on inorganic, chemically homogeneous rigid substrates (e.g., Au and $SiO_2$) through surface-interaction mechanisms[6,7]. Similar mechanisms, however, cannot achieve uniform assembly of particles on flexible polymer substrates[8,9]. The complex configurations and conformations of polymer chains create a surface with non-uniform distributions of chemical groups and phases. In addition, most assembly mechanisms require good solvent wettability, where many desirable but hard-to-wet particles and polymer substrates are excluded. Here, we demonstrate an acoustic self-limiting assembly of particles (ASAP) method to achieve wafer-scale, full-coverage, close-packed monolayers of hydrophobic particles on hydrophobic polymer substrates in aqueous solutions. We envision many applications in functional coatings and showcase its feasibility in structural coloration.**

Self-assembled monolayers (SAM) of molecules and particles show significant potentials in applications such as molecular junctions[10-12], ultrathin dielectrics[13,14], chemical and biosensors[15,16], adjustable wetting/protection coatings[17-19], and medicine[20]. SAM was originally developed for molecules based on the self-limiting chemisorption process, which involves an accurate chemical design to create the buffer layer as a bridge between the substrate and SAM-molecules[21,22]. When nano- or micro-particles are decorated with the SAM-molecules, self-limiting assembly of particles on target substrates can be achieved[23]. To achieve high-quality assembly with densely packed structures, a molecular-level chemically uniform surface of the substrate is required[24]. However, for polymers with complex configurations and conformations, it is extremely difficult to control the distribution of chemical groups or phases (e.g., amorphous and crystalline region for semicrystalline polymers)[25]. To improve the chemical uniformity, mechanical stretching has been utilized[9]. Unfortunately, high-density and low-defect assembly on flexible polymer substrates is yet to be achieved for both molecules[8,9] and molecule-decorated particles[25,26]. Although chemical treatments can be used to modify the surface properties of particles and polymer substrates, a uniform chemical treatment for polymers remains a significant challenge. In contrast to previous "chemical" strategies, our ASAP method explores a physical-process-based self-limiting strategy, utilizing interfacial energy design and viscoelastic energy dissipation in flexible polymers, where the assembly process is facilitated by acoustic and flow fields to realize high-rate, close-packed assembly of nano-/microparticles on polymer substrates.

The assembly process is outlined in Fig. 1a. In contrast to the SAM created by chemisorption of molecules with head groups interacting with the substrate[27], ASAP represents a new assembly mechanism based on physical interactions. The hydrophobic monodisperse silica ($SiO_2$) particles with a very narrow size distribution are dispersed and agitated by the acoustic field (frequency = 40 kHz, power density = 0.3 W $cm^{-2}$) in water solution to collide with a moving poly(dimethylsiloxane) (PDMS, Sylgard 184) substrate driven by a customized dipping system to provide cyclic dipping (Supplementary Fig. 1). The viscoelastic polymer substrate will attenuate the particle's kinetic energy upon collision, after which hydrophobic particles tend to stay on the hydrophobic substrate to minimize their surface energy in the water solution. Once the first particle layer forms, it becomes a hard shell on the substrate and prevents further deposition, leading to a self-limiting monolayer assembly. Such a mechanism relies on hard-to-wet surface energy design, acoustic-field-induced particle dispersion and energization, the fluid shearing at the substrate's surface, and viscoelastic polymer substrate dissipating the kinetic energy of particles. The functions of acoustic and shear fields will be discussed separately in the following section. A representative hard-to-wet assembly system is designed: de-ionized water as solvent, hydrophobic PDMS as the substrate, and hydrophobic monodisperse $SiO_2$ particles (three diameters: $d = 487±10$ nm, $761±25$ nm, and $1007±29$ nm). The characterization of particle sizes by scanning electron microscope (SEM) and atomic force microscope (AFM) can be found in Supplementary Information Fig. 2. The hydrophobicity of the particles enables them to adhere to the soft PDMS substrate due to the lower surface energy between their surface and the substrate. This is corroborated by our previous assembly work on a wide range of hydrophobic nanomaterials (e.g., graphene, h-BN, carbon black, and carbon nanotube) on hydrophobic polymer substrates (e.g., PDMS and thermoplastic polyurethan). It was demonstrated that nanomaterials with a wide size distribution can only achieve multilayer assembly[28,29]. Here, we use monodispersed $SiO_2$ particles to achieve uniform monolayer assembly. The as-received $SiO_2$ particles are hydrophilic. They are later modified with (3-aminopropyl) triethoxysilane (APTES) to create a hydrophobic surface (See surface chemistry and wettability characterization in Supplementary Figs. 3 and 4). Figs. 1b and c demonstrate a close-packed monolayer of $SiO_2$ particles ($d = 761$ nm) on a flat PDMS substrate and the coverage ($\delta$) of the assembly can reach 98.35%. Note that a patch on which particles are closely packed is considered fully covered, since no further material can fill the residual voids dictated by the particles' spherical geometry. The concentration ($C_0$) of particle suspension is 10 mg ml$^{-1}$ and the assembly time ($t$) is 900 seconds. Note, the pristine hydrophilic $SiO_2$ particles can achieve little assembly following identical procedures (Supplementary Fig. 5). The same principles can be extended to organic particles such as poly (methyl methacrylate) (PMMA) particles ($d = 5$ μm) (Supplementary Fig. 6a). Such self-limiting mechanism is not sensitive to the geometry of the substrate. We have demonstrated the conformal coating of SiO$_2$ particles ($d$ = 761 nm) on a PDMS microfiber ($d$ = 6 µm) (Fig. 1d) and PMMA particles ($d$ = 5 µm) on microtrenched PDMS (Supplementary Fig. 6b). Importantly, ASAP can be implemented within a roll-to-roll process to enable wafer-scale manufacturing as shown in Fig. 1e, Supplementary Fig. 7, and Supplementary Video 1. The assembled monolayer of SiO$_2$ particles ($d$ = 761 nm) shows a uniform orange color (Fig. 1f) when illuminated with white light from the background, highlighting its potential for structural coloration.

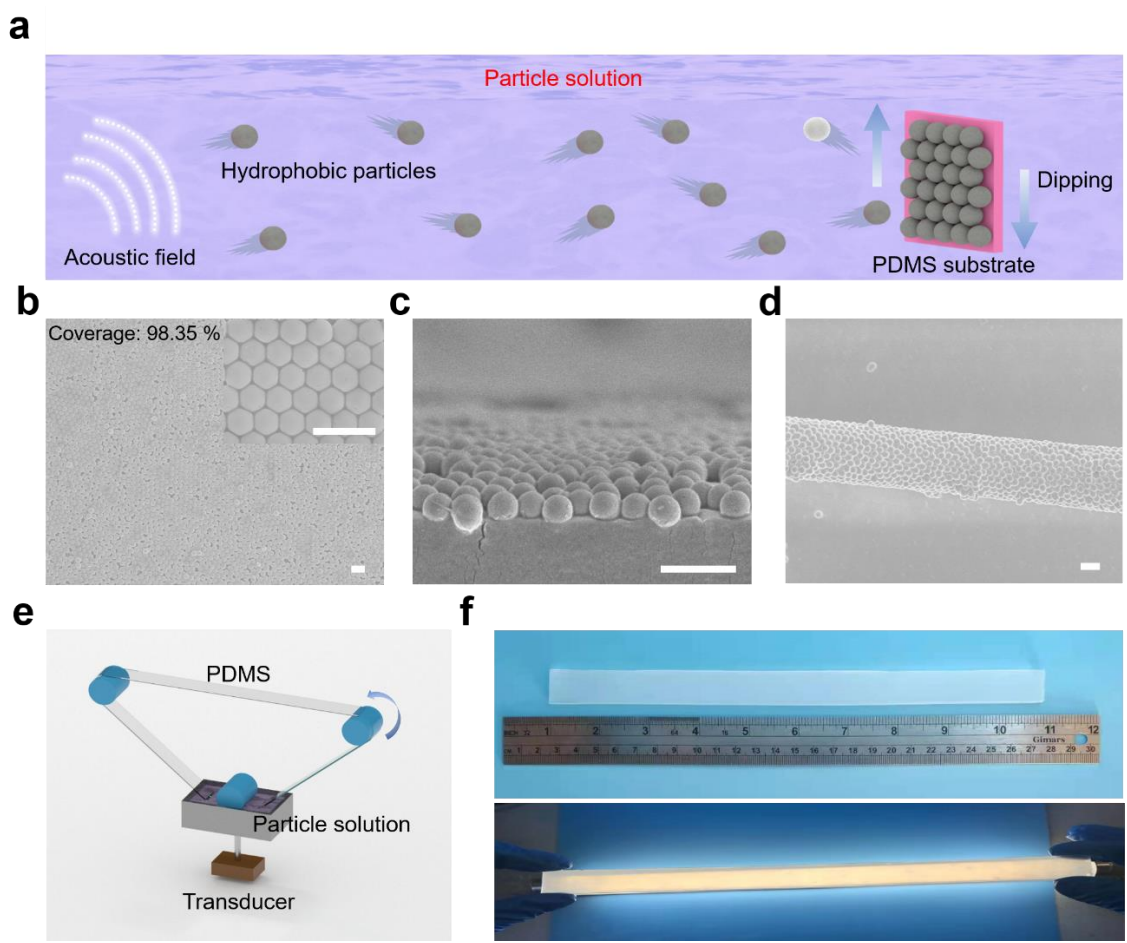

**Fig. 1 | Monolayer assembly by ASAP. a,** Schematic of ASAP process. **b,** SEM image of a monolayer of SiO$_2$ particles (1007±29 nm) on a PDMS (10:1) substrate. Insert is at

high-magnification and displays the local hexagonal structure of the particles. **c**, The cross-sectional view of monolayer $SiO_2$ particles. **d**, Monolayer $SiO_2$ particles assembled on PDMS (10:1) fiber. All scale bars, 2 μm. **e**, Schematic of large-scale manufacturing of monolayer $SiO_2$ particles on PDMS substrate. **f**, Photo of resulting large-area monolayer of $SiO_2$ particles on a PDMS substrate (top) and the produced orange color from a white light source (bottom).

Thermodynamically, the close-packed monolayer assembly takes place through two steps (Fig. 2a): random adsorption (the local minimum in the energy diagram) and reorganization (the global minimum). However, for a SAM process, the transition from step I to step II takes hours to days[9,24]. In ASAP, such a transition happens on the order of minutes. It would be intuitively anticipated that the integration of acoustic and shear fields significantly facilitates overcoming the potential energy barrier ($\Delta E$). Because both fields are applied throughout the assembly process, adsorption and reorganization happen simultaneously as demonstrated in Fig. 2b. At 5 seconds into the assembly, we observe randomly distributed particles with sporadic small aggregates. At 60 seconds, particles aggregate into bigger patches. At 120 seconds, a continuous monolayer with random voids is achieved. At 900 seconds, a close-packed monolayer is formed. Longer assembly times can further reorganize the network and reduce defects. To accurately track the assembly kinetics (coverage $\delta$ vs assembly time $t$), we fix three observation locations on the same substrate, and the SEM images taken from the three locations at the same assembly time are averaged to obtain the coverage (Supplementary Fig. 8 and 9).

To elucidate the functions of acoustic and shear fields, we fix the particle size and concentration ($d = 761\pm10$ nm and $C_0 = 10$ mg mL$^{-1}$) and compare the assembly kinetics for ASAP (Fig. 1b), assembly with acoustic field only (acoustic only), and assembly with shear field only (dipping only) (Fig. 2c-2e and Supplementary Fig. 8 and 9c). The other assembly parameters are similar among the three while we simply turn off the dipping motor (acoustic only) or acoustic transducer (dipping only). The comparison suggests that ASAP, combining both acoustic and shear fields, leads to the fastest assembly (Fig. 2c). For morphology comparison at the same assembly time ($t$ = 900 seconds), ASAP leads to a close-packed assembly (Fig. 1b), while acoustic-only process results in partial coverage (Fig. 2d) and dipping-only process leads to multilayer assembly (Fig. 2e). The acoustic effect prevents particle aggregation in the solution, and therefore keeps them from stacking up as clumps on the substrate (top image, Fig. 2d). Coverage, however, is abated by the absence of the shear that would have been provided by cyclic dipping. The shear on the substrate's surface helps the particles penetrate the thinning water film separating from the substrate during their collision with it, therefore enhancing the coverage as shown in Fig. 2e. However, a monolayer can be hardly observed (top image, Fig. 2e). Combining cyclic dipping with acoustic actuation not only leads to a monolayer covering the entire surface, but also ensures a faster assembly than one that is driven by either an acoustic field or dipping only.

Figure 2f elaborates the mechanism of ASAP on a polymer substrate. A particle heading towards the substrate at a velocity vector $\vec{V}$, consisting of a normal component, $\vec{w}$, and in-plane components $\vec{u}$ and $\vec{v}$, will induce a deformation on the flexible surface upon collision. $g(u,v,w)$ describes the particles' velocity distribution prior to surface impact while $P_d(u,v,w)$ describes the particle deposition probability. Another striking particle is much more likely to bounce off an already assembled monolayer, due to both the hard collision and the acoustic field maintaining dispersion. The time evolution of the substrate's covered fraction[30], $\delta(t) = 1 - e^{-\alpha C_0 a_p t}$, depends on the single particle's occupied area on the substrate, $a_p$, the particles' mass concentration $C_0$, and $\alpha = \frac{1}{m}\int_{-\infty}^{\infty}\int_{-\infty}^{\infty}\int_{0}^{\infty} w g(u,v,w) P_d(u,v,w) dw\, du\, dv$, where $m$ is the single particle mass. Fig.

2g shows representative fitting results of the equation for $\delta(t)$, for a particle size of 761 nm at different concentrations while Fig. 2h shows the same analysis as the one illustrated in Fig. 2g, but for different particle sizes. All coverage evolution conforms to the expected exponential trend. The authors note that higher concentrations do not necessarily ensure faster assembly and that an optimal concentration may exist. Supplemental Figures 11-14 and their associated discussion can provide some clarification, but a detailed interdependence of the parameters $\alpha$, $C_0$, and $a_p$ requires further study. We hypothesize that higher concentrations eventually become detrimental to $g(u, v, w)$ (See expression of $\alpha$), thereby slowing down the assembly process. Nevertheless, the ASAP method achieves monolayer assembly regardless of particle concentration and size within the testing range.

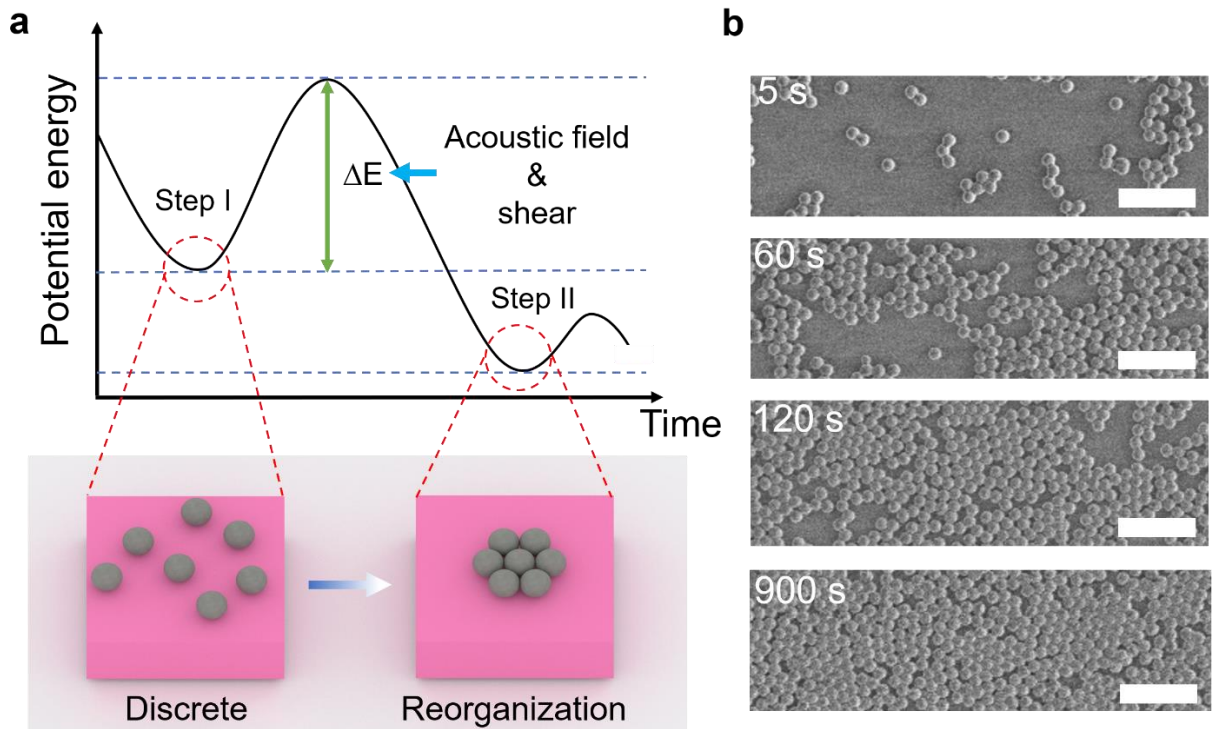
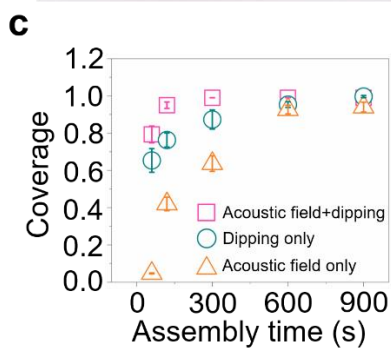
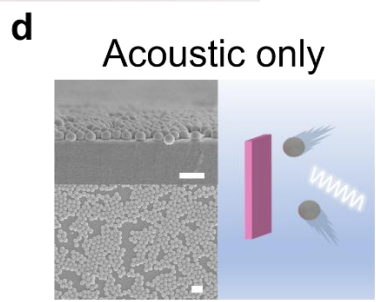
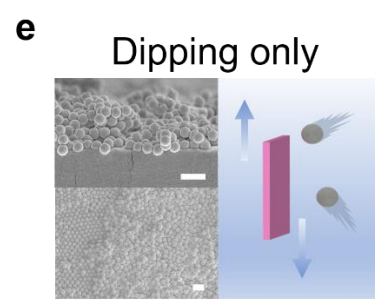
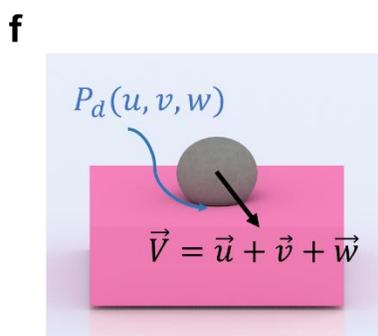
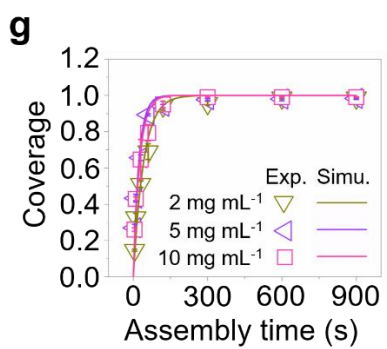
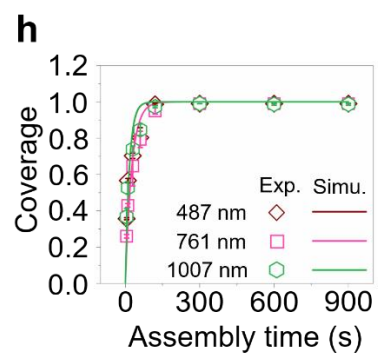

**Fig. 2 | Self-limiting assembly monolayer mechanism from thermodynamic and kinetic perspectives. a,** Potential energy evolution with respect to assembly time and schematic of particle reorganization during the assembly process. $\Delta E$ represents the potential energy barrier. **b**, SEM images of particle reorganization. **b**, Scale bars, 5 µm. **c**, Coverage evolution using acoustic field + dipping, acoustic only, and dipping only. **d,e**, Planar and cross-sectional SEM images of SiO$_2$ particles onto PDMS substrate (761 nm, 10 mg mL$^{-1}$) after 15 min assembly using different methods and their corresponding assembly schematics. **d,** Acoustic only**. e,** Dipping only. All scale bars in **d** and **e**, 2 µm. **f**, A diagram of the collision and adhesion process of a single particle on polymer substrate. **g**, Experiment and simulation of coverage evolution at different concentrations of SiO$_2$ (761 nm in diameter) and **h**, same concentration (10 mg mL$^{-1}$) with different SiO$_2$ diameters.

One potential application of producing large-scale self-assembled self-limiting films of micro/nanoparticles is structural coloration. In contrast to conventional coloration methods using organic molecules in pigments that are facing the challenge of instability and toxicity, colors from engineered architectures that are originally inspired by natural creatures offers promising alternative routes[31-34]. Particularly, dielectric structures enable high-quality optical resonance by reducing the materials loss that is inevitable in metallic nanostructures[35]. Furthermore, creating 3D structures delivers more design freedom to cover a full range spectrum of colors[36]. However, the top-down lithography-based manufacturing is time-consuming, sophisticated, and cost-inefficient. Here, as a proof of concept, we used the obtained monolayer and stacked multiple monolayers of self-assembled SiO$_2$ particles to create different structure-induced colors recorded using our lab-made observing apparatus (Supplementary Fig. 15).

Figure 3a summarizes the colors observed in the monolayer films of self-assembled SiO$_2$ particles manufactured by ASAP with different particle diameters; 487 nm diameter particles display a yellow color, 761-nm-diameter particles display an orange color, and 1007 nm diameter particles display a blue color. The corresponding attenuation spectrum for 1007-nm-diameter particles is shown in Fig. 3b, where a clear resonance feature is observed. Fig. 3b displays excellent agreement between finite-difference-time-domain (FDTD) simulation and experimental scattering spectra of our $SiO_2$ (1007 nm) monolayer particles. The scattering spectra of the other $SiO_2$ monolayer particles with different diameters can be found in Supplementary Fig. 17. Furthermore, the FDTD-simulated Ez-field profile in x-y plane shown in Fig. 3c when the electromagnetic wave propagates along z-axis suggests that the structural coloration derives from the magnetic dipole Mie resonance in the visible range. According to Mie theory[37], the resonance peak depends on the refractive index of materials and scatterer dimensions. In Supplementary Figs. 16a and b, it is clear that the resonance peak shifts to red with an increasing diameter of nanospheres and refractive index of materials. During the ASAP assembly process, the particles are accelerated by the acoustic wave and slammed onto the substrate. Considering the viscoelastic elasticity of PDMS substrates, the particles would get partially embedded inside the substrate (Supplementary Fig. 18). As the refractive index of $SiO_2$ is close to PDMS, the embedment reduces the effective scattering volume of particles. Thus, we calculate the depth of $SiO_2$ particles ($D_{dep}$) inside the PDMS substrate (Supplementary Fig. 18). As demonstrated in Supplementary Fig. 16c, the simulated Mie scattering spectra as a function of $D_{dep}$ indicate that larger $D_{dep}$ causes blue shifting of the resonance peak.

Based on the parameters that can influence the transmission spectra mentioned above, we measured those parameters by AFM and SEM (Supplementary Fig. 17) and did the calculations of converting measured and FDTD-simulated transmission spectra to colors. From Fig. 3d, the calculation results display excellent agreement with experimental structural coloration under white light illuminiation. Furthermore, we calculate colors for a group of particles with different diameters and plot their color coordinates in CIE 1931 color space; see Fig. 3e. The results show the color range of monolayer films of particles whose diameters range from 200 nm to 1500 nm. In addition, we stacked the monolayer samples (Fig. 3f), and experimental colors of different combinations are summarized in Fig. 3g. The results show that the stacked monolayers can significantly enrich the color map and the potential to create a full range of coloration.

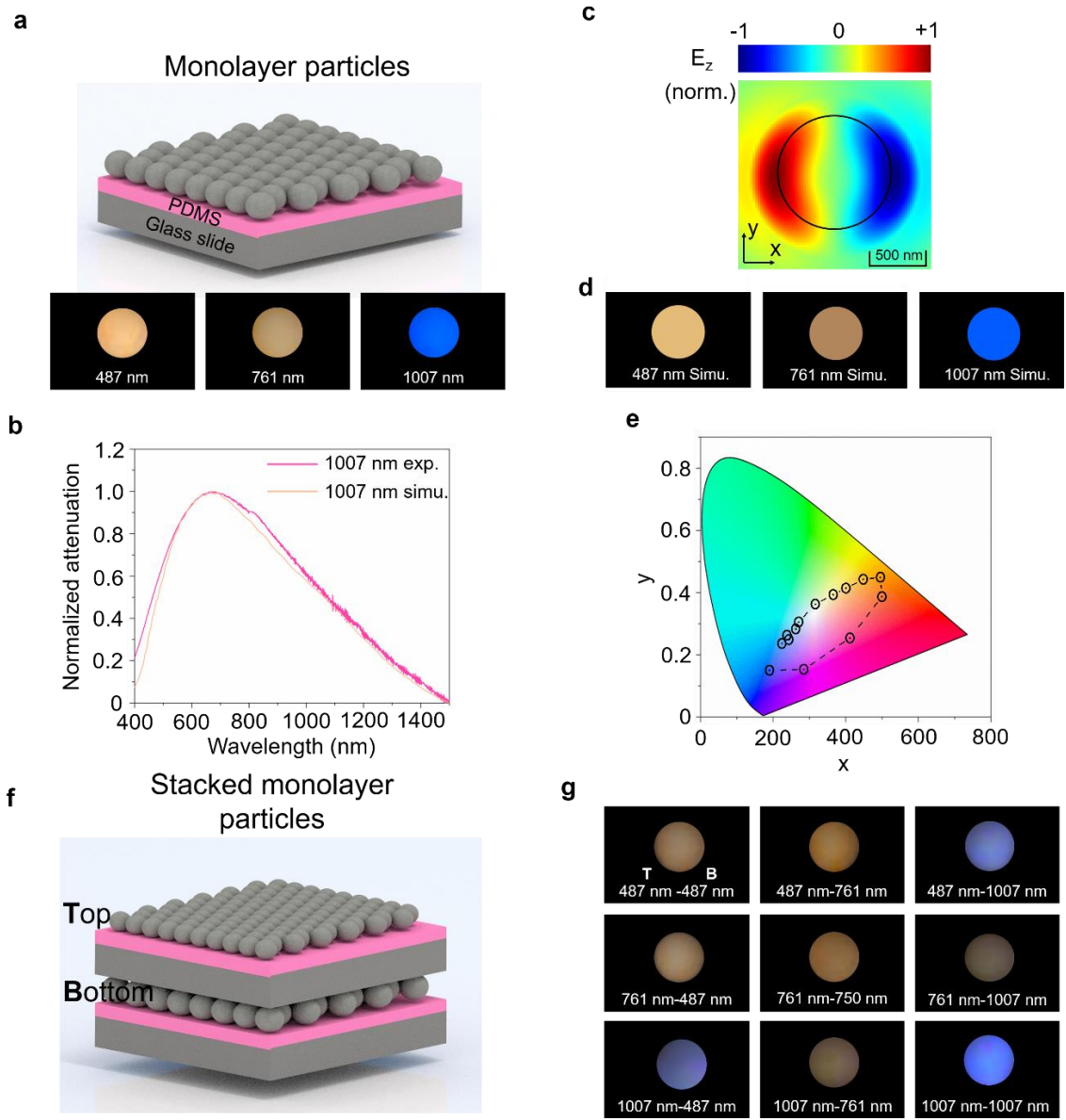

**Fig. 3 | Structural coloration of monolayer SiO$_2$ assembled by ASAP. a**, Schematic of assembled monolayer SiO$_2$ on PDMS coated glass slide and structural coloration with respect to various SiO$_2$ diameters. **b**, Experimental and simulation scattering spectra of monolayer SiO$_2$ (1007 nm in diameter) particles. **c**, The simulation electric field distribution of magnetic dipole (MD) mode at 0.66 $\mu$m. **d**, Simulation color images based on measured sample parameters. **e**, CIE1931 chromaticity diagram and simulation color space values obtained from scattering spectra of SiO$_2$ nanosphere. **f**, Schematic of stacked monolayer SiO$_2$ particles. **g**, Structural coloration based on stacked monolayer SiO$_2$ particles and number pair indicates the stacking order where the left number

indicates the diameter of the particle on the bottom layer and right number indicates the diameter of the particle on the top layer. Note, here B and T represent the bottom layer and top layer with different or same particle sizes according to the stacking sequence.

In summary, ASAP represents a high-quality and scalable self-limiting assembly method to achieve monolayer particles on flexible polymer substrates. Unlike chemisorption mechanisms for SAM, ASAP highlights a new assembly mechanism in which physical collision and surface adhesion between particle and viscoelastic polymer substrates are modulated by acoustic and flow fields. Such new mechanism is independent of the chemical heterogeneity of polymer substrates and enables close-packed monolayer particles on polymer. ASAP also unlocks a wide range of hard-to-wet systems that are desirable but extremely challenging to manufacture. The capability of accurately controlling the layer numbers of assemblies in wafer scale with full coverage paves the way towards next-generation flexible optics and wearable electronics.

**Author contributions**
B. L. conceived, initiated, and supervised the project. L. Z. conceived the experiment and performed the assembly, structural characterization including all SEM imaging, contact angle measurement and part of AFM measurement, and structural coloration characterization. B. S. performed the coverage of assembly and analysis and simulation of the assembly data. J. F. performed the analysis and simulation of structural coloration. R. C. performed the measurement of scattering spectra. T. S. constructed the roll-to-roll set-up and performed part of the assembly. S. D. performed part of AFM measurement. W. G. designed and supervised the optical measurement and data analysis. Q. W. supervised physics-based modeling. All authors contributed to writing and revising the manuscript.

**Acknowledgments**
L.Z., B.S., Q.W. and B.L. were partially supported by the National Science Foundation CMMI Advanced Manufacturing Program under Award # 2003077. L.Z., and B.L. acknowledge the startup fund of Villanova University. J. F., R. C., and W. G. acknowledge the startup fund of the University of Utah.

**Competing interests**

B. L., B. S. and Q. W. have a U.S. Provisional Patent Application on the topic of this paper (**63/135,872**, 2021). The other authors declare no competing interests.

# Supplementary Information

# Wafer-scale, full-coverage, acoustic self-limiting assembly of particles on flexible substrates

Liang Zhao[1,2], Bchara Sidnawi[1,3#], Jichao Fan[4#], Ruiyang Chen[4], Thomas Scully[1,2], Scott Dietrich[5], Weilu Gao[4], Qianhong Wu[1,3], Bo Li[1,2,*]

[1] Department of Mechanical Engineering, Villanova University, Villanova, PA 19085, USA.

[2] Hybrid Nano-Architectures and Advanced Manufacturing Laboratory, Villanova University, Villanova, PA, 19085, USA

[3] Cellular Biomechanics and Sports Science Laboratory, Villanova University, Villanova, PA, 19085, USA

[4] Department of Electrical and Computer Engineering, University of Utah, Salt Lake City UT, 84112, USA

[5] Department of Physics, Villanova University, Villanova, PA, 19085, USA

[#] These authors contributed equally to this work.

* Corresponding author: bo.li@villanova.edu

## Methods:

**Modification of SiO₂ particles**

Three kinds of monodisperse silicon dioxide (SiO$_2$) spheres (Sigma; 500 nm, 750 nm, 1000 nm) are employed in this work. The diameters are further precisely determined by SEM images. The results show that the average diameters are 487±10 nm, 761±25 nm, and 1007±29 nm, respectively. Because of the intrinsic hydrophilicity of SiO$_2$, we use (3-aminopropyl)triethoxysilane to modify it and transfer it into hydrophobicity. Briefly, 100 mg SiO$_2$ particles are added into 10 mL ethanol, followed by sonication for 5 min. Then the solution is heated in an oil bath at 60 °C for 48 h. The resulted solution finally is centrifuged for 5 times (each time for 5 min).

**Self-limiting assembly of monolayer**

Polydimethylsiloxane film (PDMS) is fabricated on the surface of glass slide (1 inch × 1 inch) using spin-coating method. The ratio of monomer and cross-linking agent is adjusted to 10:1 in volume. Note, this is for the assembly of SiO$_2$ particles, but 20:1 for PMMA particle assembly. The PDMS fiber is prepared by stretching the uncured PDMS on a glass slide (heated to 100 ℃). The patterned PDMS is made in a patterned Si wafer mold. Before assembly, the as-prepared hydrophobic particle solution is heated to 50 °C and kept in a sonicator. The spin-coated PDMS substrate then is held in the holder of a lab-made dip coater. During the dipping process, the sonication is applied, and the sample is immersed in the solution all the time.

**Structural coloration simulation**

To reveal the origin of the structural coloration produced by our ASAP sample, we employed finite-difference time-domain method to fit experimental spectra of monolayer nanoparticles. To accelerate the simulation, we employ periodical boundary conditions because of the lattice structure in our samples, in which the refractive indexes of $SiO_2$ and PDMS are set to be constant values, as $n_{SiO_2} = 1.45$ and $n_{PDMS} = 1.43$, respectively. A total-field scattered-field source is utilized to separate scattered field from total field, so that we can measure the scattering spectrum of particles that are illuminated by a plane wave.

The spectra are first converted to XYZ color coordinates using Eqs.1-3:

$$X = \int R(\lambda)\bar{x}(\lambda)d\lambda, \tag{1}$$

$$Y = \int R(\lambda)\bar{y}(\lambda)d\lambda, \tag{2}$$

$$Z = \int R(\lambda)\bar{z}(\lambda)d\lambda, \tag{3}$$

Where $R(\lambda)$ is the scattering spectra, the color matching functions $\bar{x}(\lambda)$, $\bar{y}(\lambda)$, $\bar{z}(\lambda)$ defined by International Commission on Illumination (CIE) are a numerical description of the response of the three types of cone cells in the human eye. After normalization using Eqs. 2

$$x = \frac{X}{X+Y+Z}, \quad y = \frac{Y}{X+Y+Z}, \quad z = \frac{Z}{X+Y+Z}, \tag{2}$$

only two parameters, $x$ and $y$, are needed to locate the color position in CIE1931 color space and convert the spectrum to color.

**Instrumentation**

Scanning electron microscope (SEM, Hitachi S-4800) is used to characterize the sample morphologies. The wettability is conducted using a lab-made contact angle measurement set-up. The DI water droplet is ~5 µL. The Fourier transform infrared (FTIR, PerkinElmer) spectrum is conducted to character the surface chemistry of $SiO_2$ particles. The $SiO_2$ particle diameter and depth into PDMS are obtained by atomic force microscopy (AFM, Agilent 5500). The scattering spectra of monolayer $SiO_2$ particles are conducted by Hitachi U4100 UV-Vis-NIR spectrometer.

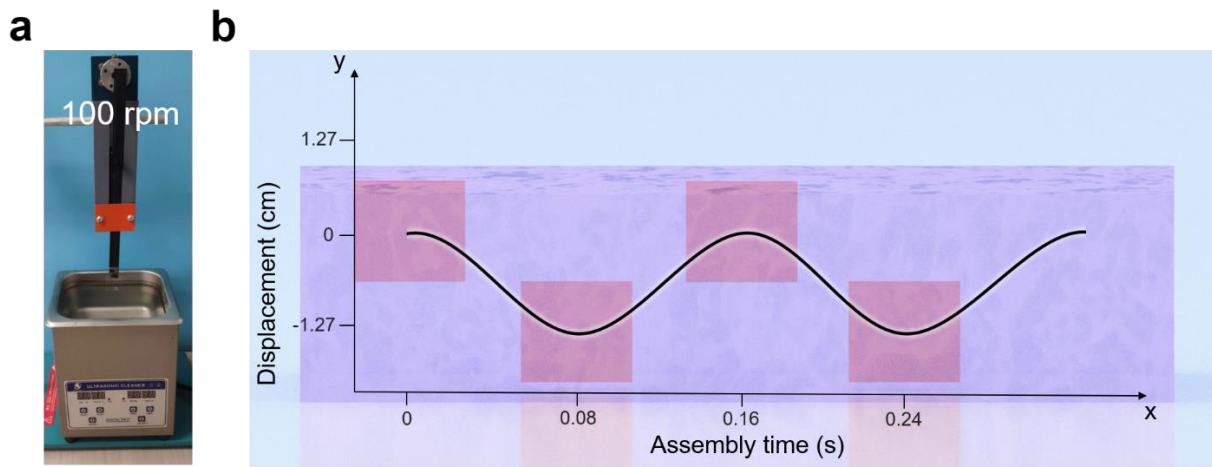

**Supplementary Fig. 1 | The set-up of sono + dipping process. a,** Digital image. **b,** Displacement of the PDMS substrate (pink square) as a function of assembly time. The purple background represents the solution, and the substrate is submerged in the solution during the entire dipping process.

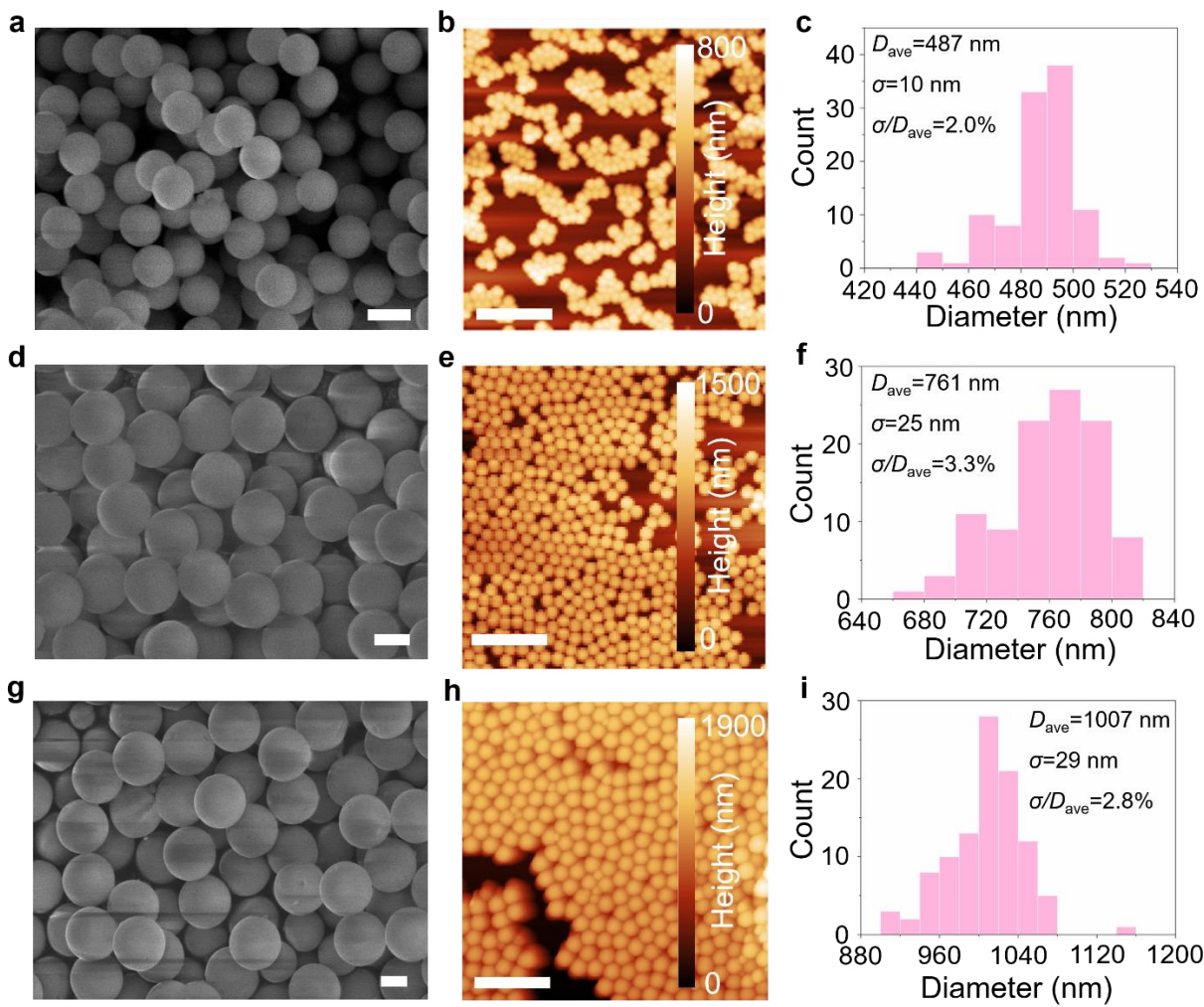

**Supplementary Fig. 2 | SEM images, AFM images, and diameter distribution of monodisperse SiO₂ spheres**. **a-c**, 487 nm. **d-f**, 761 nm. **g-i**, 1007 nm. a, d, g, Scale bars, 500 nm. It should be noted that the AFM images show the monolayer SiO₂ spheres on Si wafer, of which assembly is induced by capillary force. The average diameters ($D_{ave}$) are calculated from AFM images by measuring Z-heights (more than 100 spheres). Scale bars, 5 μm.

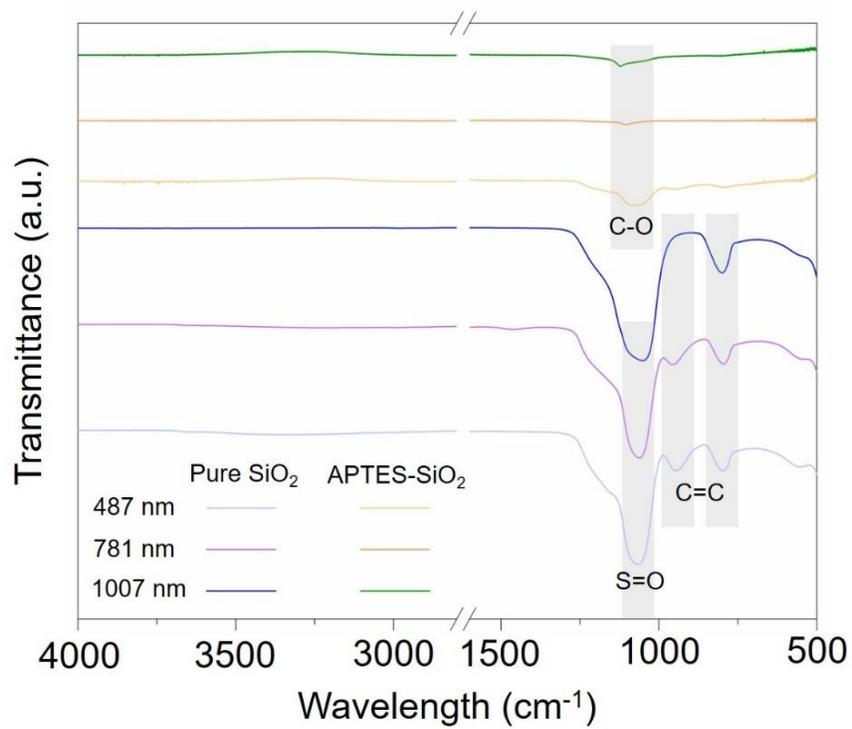

**Supplementary Fig. 3 | FT-IR spectra of pure and APTES modified SiO$_2$ (APTES-SiO$_2$) spheres.**

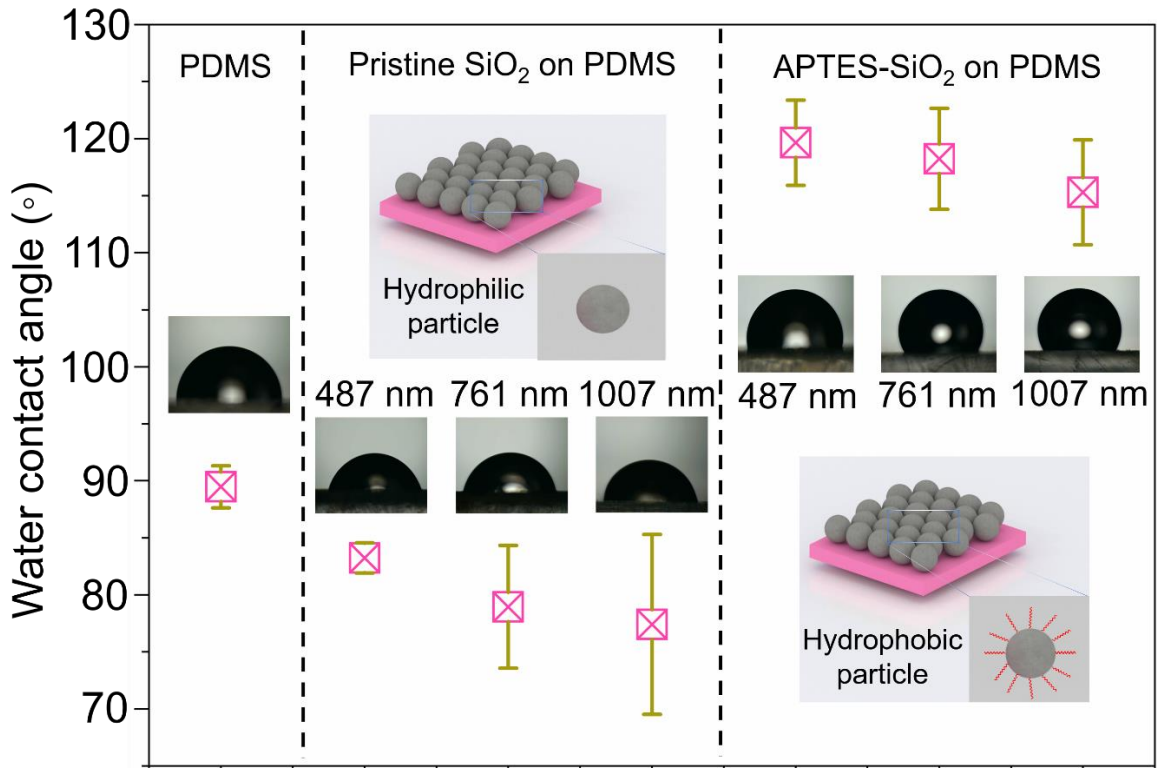

**Supplementary Fig. 4 | Contact angle measurement.** The pristine SiO$_2$ particles (i.e., hydrophilic) are rubbed onto PDMS substrate. The APTES-SiO$_2$ particles (i.e., hydrophobic) are assembled on PDMS substrate by ASAP method.

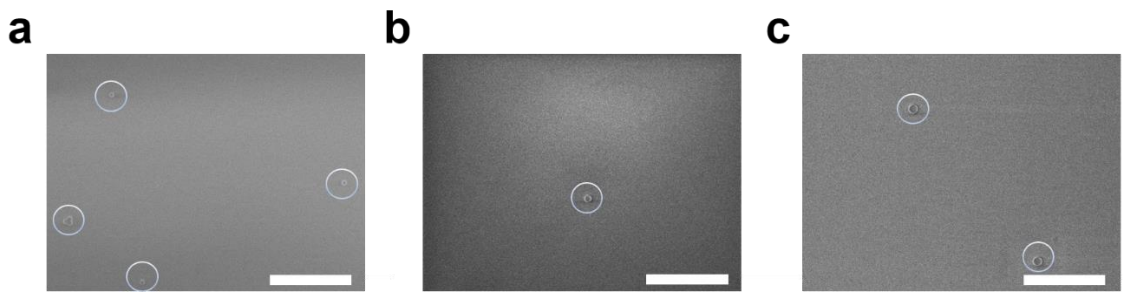

**Supplementary Fig. 5 | Pure SiO₂ particles (i.e., hydrophilic) assembled on PDMS substrate (10:1) using ASAP. a**, 487 nm-10 mg mL$^{-1}$. **b**, 761 nm-10 mg mL$^{-1}$. **c**, 1007 nm-10 mg mL$^{-1}$. Scale bars, 50 µm. Here, we used the pure SiO₂ particles as received to do the assembly (assembly time: 15 min). The SEM images reveal that hydrophilic SiO₂ almost can't be assembled using ASAP (partial assembly in marked region), which is ascribed from the high restraining of water to hydrophilic particles. That means the hydrophilic particles can't escape from the water and be assembled on the hydrophobic substrate.

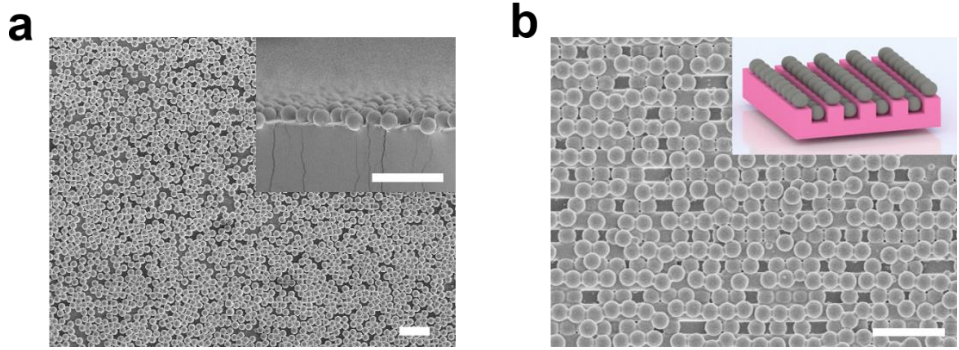

**Supplementary Fig. 6 | PMMA (~5 μm) particles assembled on PDMS substrate (20:1). a,** Flat PDMS substrate. Insert is the cross-sectional view of monolayer $SiO_2$. **b,** Microtrenched PDMS. Insert is the schematic of patterned PDMS with assembled PMMA particles. Scale bars, 20 μm.

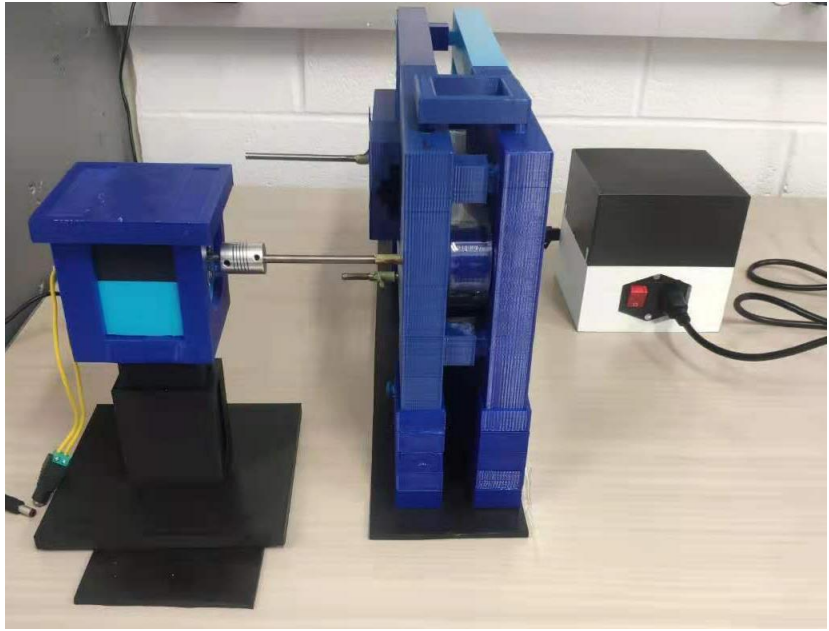

**Supplementary Fig. 7 | The set-up of roll-to-roll apparatus.**

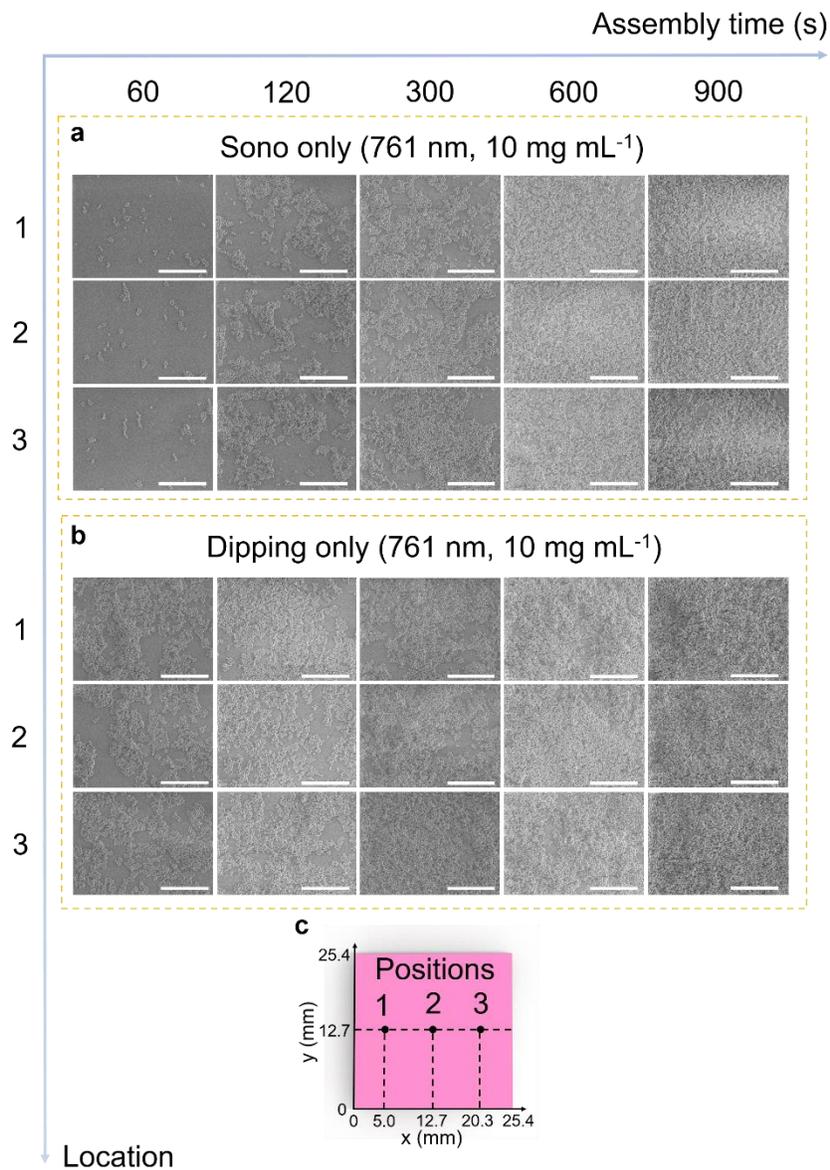

**Supplementary Fig. 8 | SEM images of assembly process of SiO$_2$ particles (761 nm-10 mg mL$^{-1}$) using different assembly procedures. a**, Acoustic only. **b**, Dipping only. Scale bars, 20 μm. **c**, Selected positions on samples for SEM imaging. To track the evolution of the assembly, once the locations are chosen, the SEM stage will navigate to the same locations of the same substrates for imaging. Note, all the evolutions for assembly in this paper follow the positions shown in **c.**

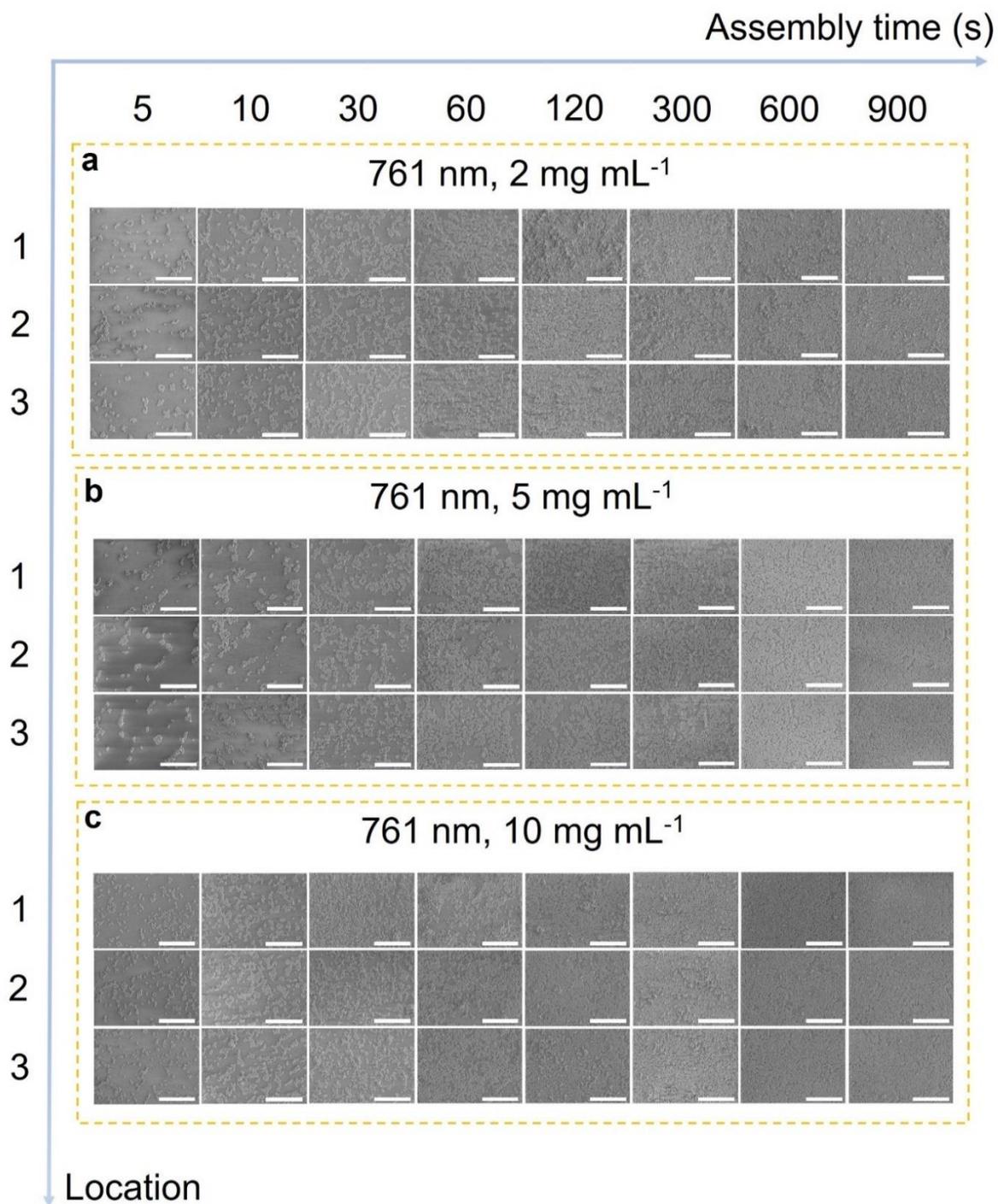

**Supplementary Fig. 9 | SEM images of assembly process of SiO$_2$ nanoparticles (761 nm) with various concentrations**. **a**, 2 mg mL$^{-1}$. **b**, 5 mg mL$^{-1}$. **c**, 10 mg mL$^{-1}$. Scale bars, 20 μm.

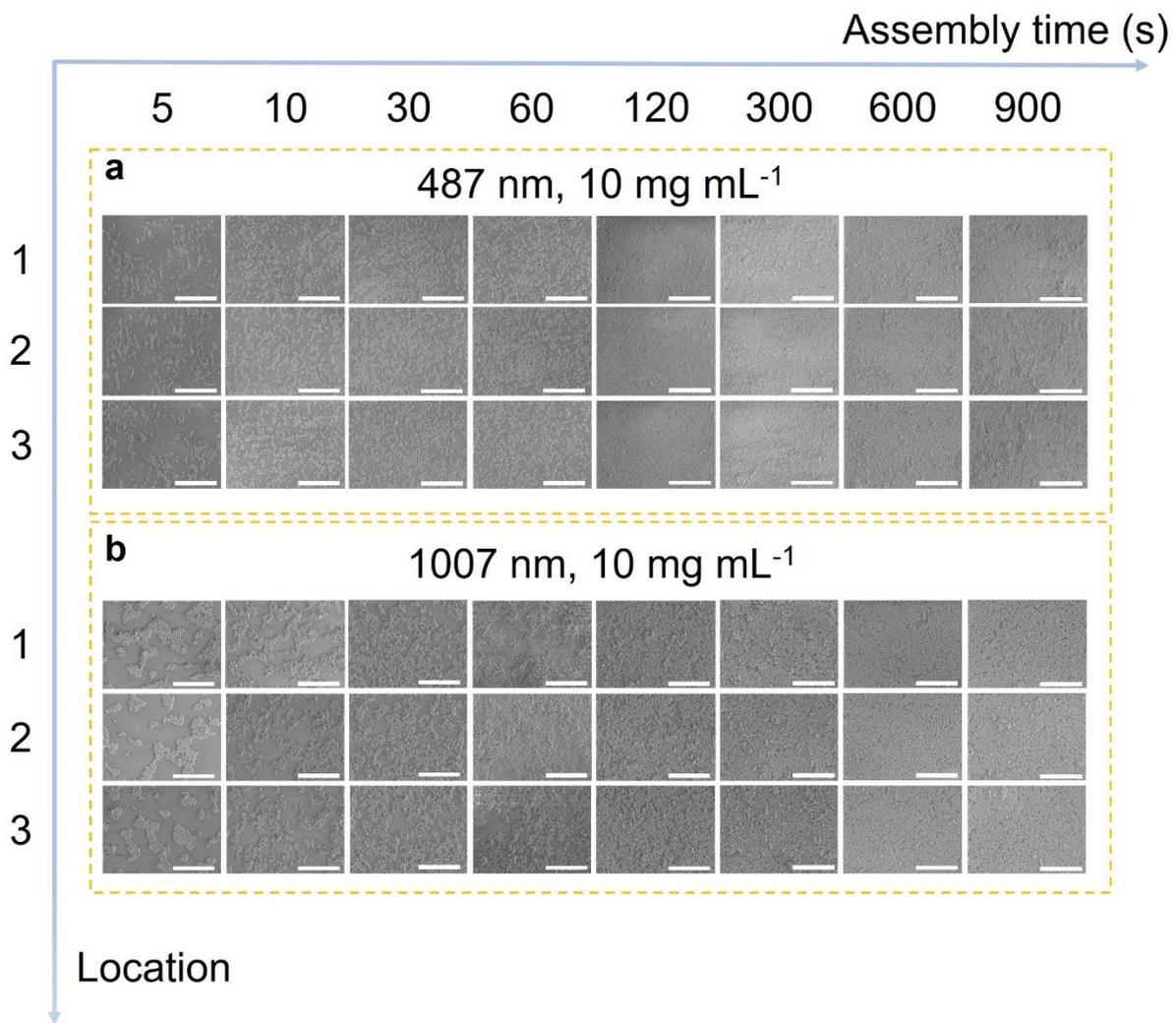

**Supplementary Fig. 10 | SEM images of assembly process of SiO$_2$ nanoparticles (10 mg mL$^{-1}$) with different diameters**. **a**, 487 nm. **b**, 1007 nm. Scale bars, 20 μm.

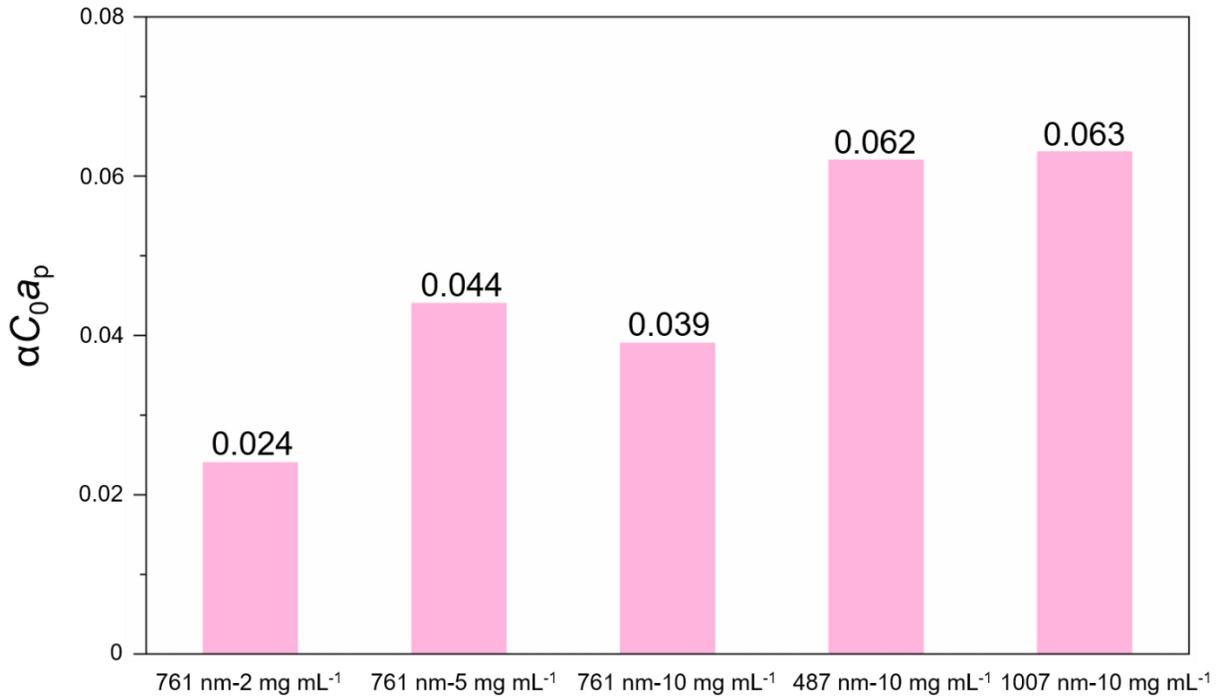

**Supplementary Fig. 11 | Summary of values of $\alpha C_0 a_p$ for assembly process using various SiO₂ nanoparticle diameters and concentrations.**

Since the particle size is the same for all 3 cases, $a_p$ is also the same. Therefore, if the value of $\alpha$ were to be unchanged, $\alpha C_0 a_p$ must be proportional to $C_0$. However, one finds that the value of $\alpha C_0 a_p$ first increases with the increase of $C_0$ from 2 to 5 mg mL$^{-1}$, but then decreases after $C_0$ further increases to 10 mg mL$^{-1}$, suggesting that more particles do not necessarily ensure faster coverage. It is a clear indication that the value of $\alpha$ is decreasing with increasing concentration. The value of $\alpha$ cannot be affected by $m$ or $P_d$ since those are determined by the particle size, and the particle-substrate interfacial interactions, respectively, which are the same for all 3 cases. The particle velocity distribution, $g(u, v, w)$, is determined by the dipping regimen, and the particles interaction with the acoustic field. Since the former is identical for all cases, the latter seems to be the main underlying cause of the behavior exhibited by $\alpha C_0 a_p$. Because the acoustic power and frequency settings were the same for the 3 cases, a quite plausible explanation for this behavior is the increased dispersion of acoustic energy at higher concentrations. Therefore, for a given particle size, acoustic settings, and dipping regimen, an optimal concentration that elicits the fastest assembly, exists. Fig. 2h shows the intermediate particle size of 761 nm leads to the poorest coverage performance, compared to those corresponding to the other two sizes. Systematic control of vectorial acoustic field in future experiments will offer more insights into revealing the effect of particle size. In addition to $a_p$ being different, different particle sizes imply different values of $m$ and $P_d$ in the expression of $\alpha$. Also, it is worth noting that the particle size relative to the wavelength of the sound waves is crucial to its interaction with the acoustic field, which affects $g(u, v, w)$. Therefore, future experiments, in which the acoustic field is systematically controlled, where frequencies may be such that wavelengths are of comparable size to the particle's, are essential to achieve a clear understanding of the size effect.

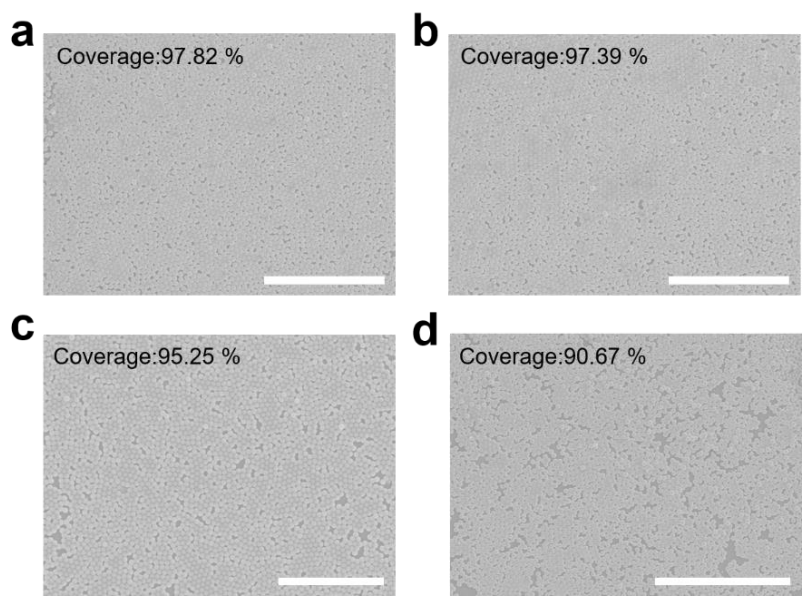

**Supplementary Fig. 12 | Plane SEM images of assembled monolayer SiO$_2$ with various concentrations and diameters. a**, 2 mg mL$^{-1}$-761 nm. **b**, 5 mg mL$^{-1}$-761 nm. **a, b**, Scale bars, 20 μm. **c**, 10 mg mL$^{-1}$-487 nm. Scale bar, 10 μm. **d,** 10 mg mL$^{-1}$-1007 nm. Scale bar, 30 μm.

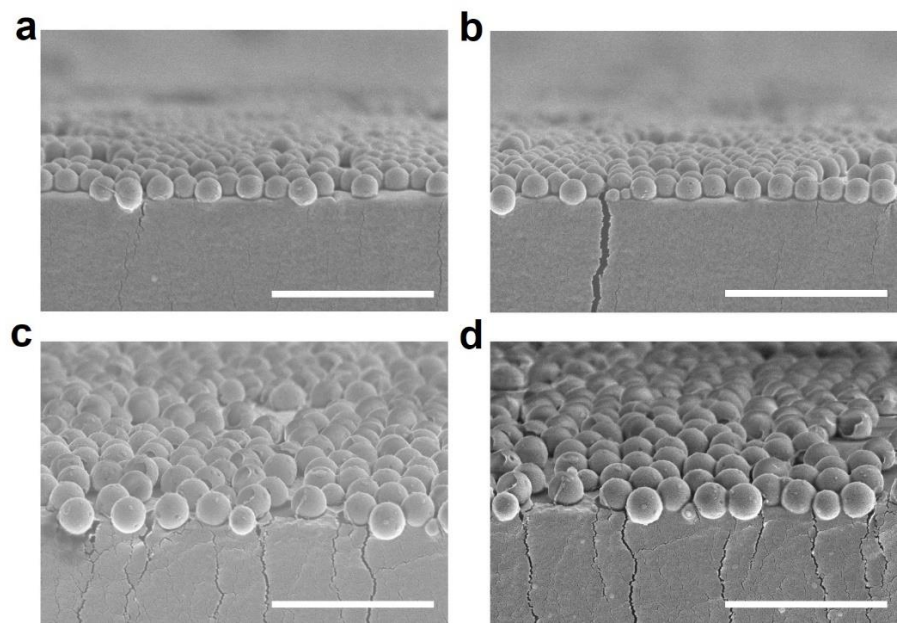

**Supplementary Fig. 13 | Cross-sectional SEM images of assembled monolayer SiO$_2$ with various concentrations and diameters. a**, 2 mg mL$^{-1}$-761 nm. **b**, 5 mg mL$^{-1}$-761 nm. **c**, 10 mg mL$^{-1}$-487 nm. **d**, 10 mg mL$^{-1}$-1007 nm. Scale bars, 5 μm.

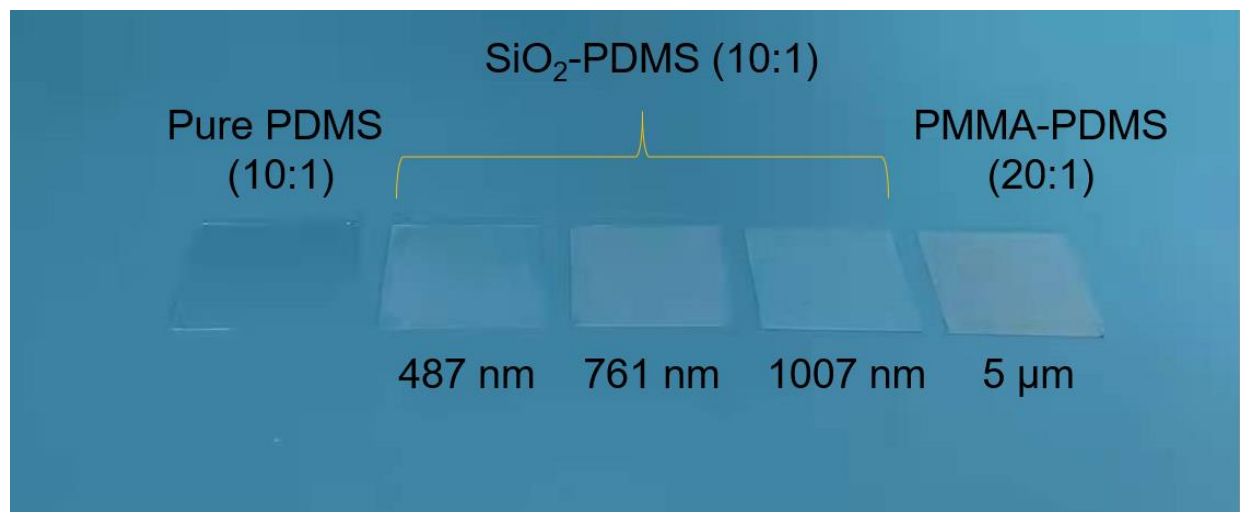

**Supplementary Fig. 14 Digital images of assembled monolayer particles on PDMS coated glass slide. a,** Pure PDMS. **b,** 487 nm $SiO_2$. **c,** 761 nm $SiO_2$. **d,** 1007 nm $SiO_2$. **e, 5 μm** PMMA spheres.

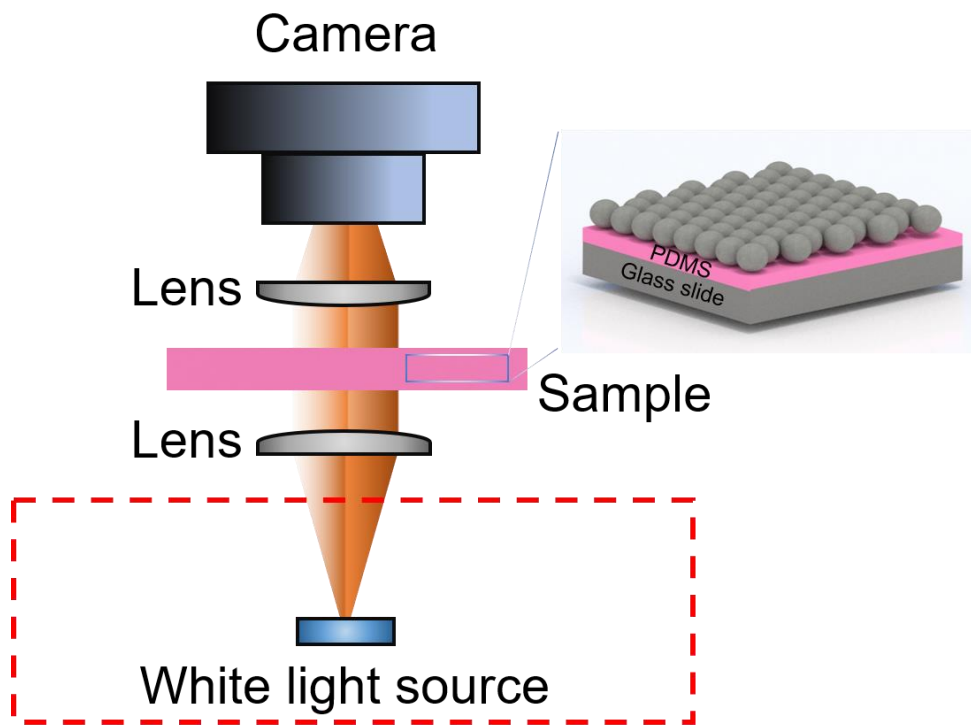

**Supplementary Fig. 15 | The set-up of observation of structural coloration.** The white light source is provided by a LED light (2.5 V), which is located at a closed chamber. In the center of top of the chamber, there is a spherical hole allowing the white light to shoot outside. Then the light will go through lens-sample-lens, and finally the camera will take the image for structural coloration. It should be noted that the lens is to make the white perpendicular to the sample so as to reduce the halo.

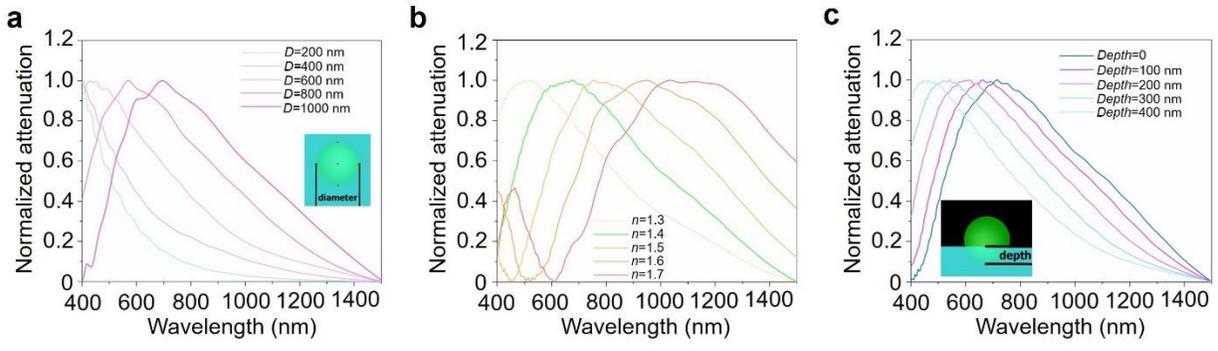

**Supplementary Fig. 16 | Scattering spectra simulation of monolayer SiO$_2$ nanoparticles using various parameters. a**, Particle diameters. **b**, Refractive index of particles. **c**, The embedded depth of SiO$_2$ particles into PDMS substrate.

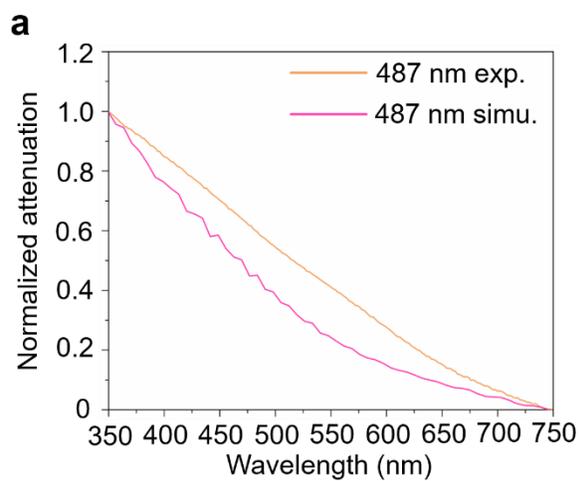 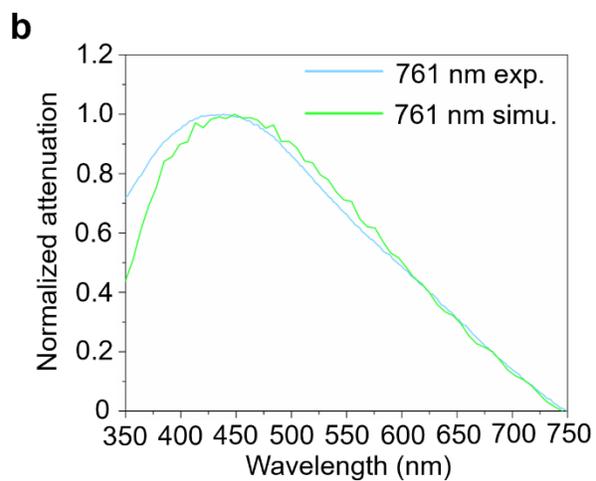

**Supplementary Fig. 17 | Experimental and simulation scattering spectra of monolayer SiO$_2$ particles. a**, 487 nm. **b**, 761 nm.

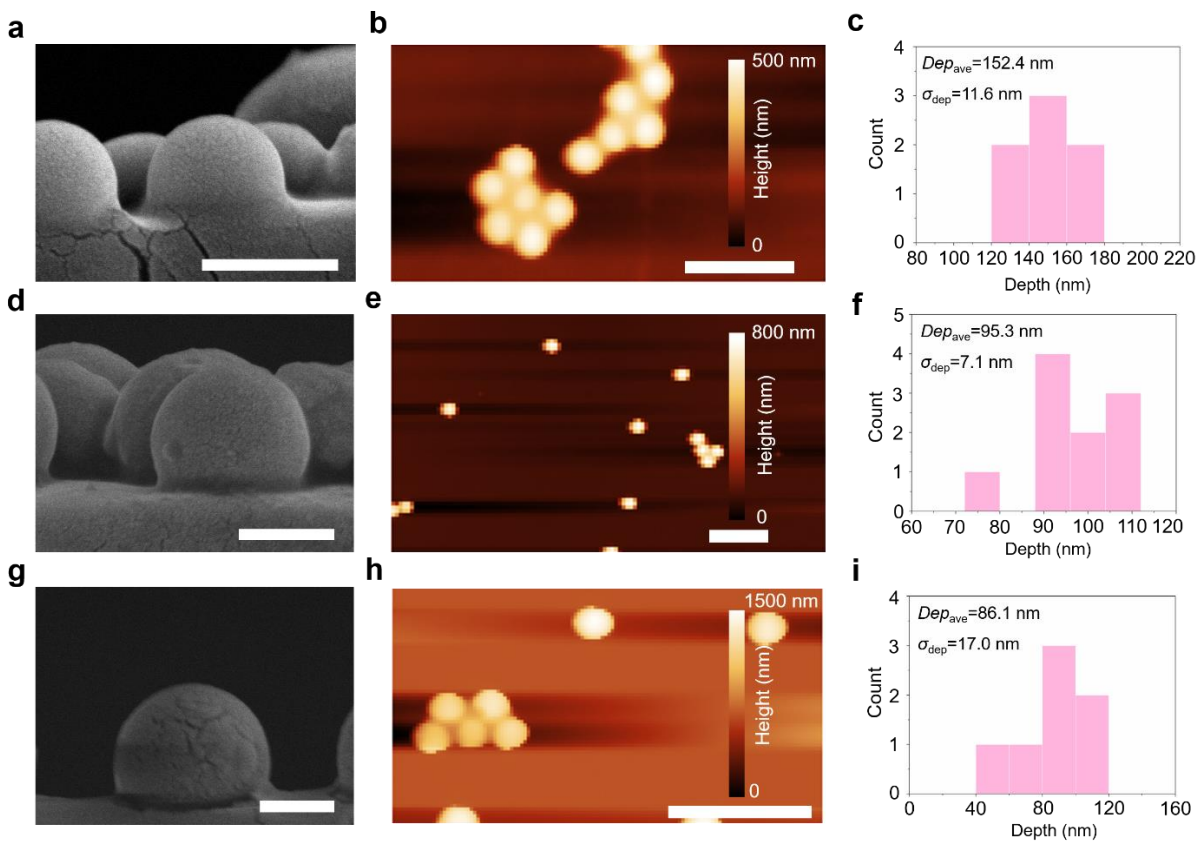

**Supplementary Fig. 18 Depth of SiO₂ particles into PDMS (10:1) substrate.** SEM images, AFM images, and calculated depth SiO₂ particles on the PDMS (10:1) substrate. **a-c**, 487 nm. **d-f**, 761 nm. **g-i**, 1007 nm. **a, d, g**, Scale bars, 500 nm. **b**, e, h, Scale bars, 10 µm.